\documentclass[aps,prl,twocolumn,showpacs,superscriptaddress]{revtex4-1}

\usepackage[english]{babel}
\usepackage{graphicx}
\usepackage{dcolumn}
\usepackage{amsmath}
\usepackage{epsfig}
\usepackage{bm}
\usepackage{amssymb}
\usepackage{float}
\usepackage{natbib}
\usepackage{color}
\usepackage{soul}
\usepackage{hyperref}
\usepackage{lineno}
\hypersetup{colorlinks=true,linkcolor=blue,citecolor=blue,urlcolor=black}

\usepackage{units}

\usepackage{hyperref}
\hypersetup{%
colorlinks,
linkcolor={black},
citecolor={black},
urlcolor={black}}

\newcommand{\change}[1]{\color{black} #1 \color{black}}
\newcommand{\comment}[1]{\color{black}\it \color{black} \rm}
\newcommand{\changes}[1]{\color{black} #1 \color{black}}

\newcommand{\ket}[1]{| #1 \rangle}
\newcommand{\bra}[1]{\langle #1 |}

\begin{document}

\title{Optical coherent manipulation of alkaline-earth circular Rydberg states}

 \author{Andrea Muni$^*$}
\affiliation{Laboratoire Kastler Brossel, Coll\`ege de France,
 CNRS, ENS-Universit\'e PSL,
 Sorbonne Universit\'e, \\11, place Marcelin Berthelot, 75005 Paris, France}

 \author{L\'ea Lachaud$^*$}
\affiliation{Laboratoire Kastler Brossel, Coll\`ege de France,
 CNRS, ENS-Universit\'e PSL,
 Sorbonne Universit\'e, \\11, place Marcelin Berthelot, 75005 Paris, France}
  \email{These two authors contributed equally.}
  
 \author{Angelo Couto}
\affiliation{Laboratoire Kastler Brossel, Coll\`ege de France,
 CNRS, ENS-Universit\'e PSL,
 Sorbonne Universit\'e, \\11, place Marcelin Berthelot, 75005 Paris, France}

 \author{Michel Poirier}
\affiliation{CEA Universit\'e Paris-Saclay, IRAMIS,  Laboratoire ``Interactions, Dynamiques et Lasers'', 91191 Gif sur Yvette, France}

\author{Raul Celistrino Teixeira}
\affiliation{Laboratoire Kastler Brossel, Coll\`ege de France,
 CNRS, ENS-Universit\'e PSL,
 Sorbonne Universit\'e, \\11, place Marcelin Berthelot, 75005 Paris, France}
 \affiliation{Departamento de F\'isica, Universidade Federal de S\~ao Carlos, Rod. Washington Lu\'is, km 235 - SP-310, 13565-905 S\~ao Carlos, SP, Brazil}
 
 \author{Jean-Michel Raimond}
\affiliation{Laboratoire Kastler Brossel, Coll\`ege de France,
 CNRS, ENS-Universit\'e PSL,
 Sorbonne Universit\'e, \\11, place Marcelin Berthelot, 75005 Paris, France}

\author{Michel Brune}
\affiliation{Laboratoire Kastler Brossel, Coll\`ege de France,
 CNRS, ENS-Universit\'e PSL,
 Sorbonne Universit\'e, \\11, place Marcelin Berthelot, 75005 Paris, France}

 \author{S\'ebastien Gleyzes}
 \affiliation{Laboratoire Kastler Brossel, Coll\`ege de France,
 CNRS, ENS-Universit\'e PSL,
 Sorbonne Universit\'e, \\11, place Marcelin Berthelot, 75005 Paris, France}

\hyphenation{Ryd-berg sen-sing ma-ni-fold}

\maketitle

\date{\today}

\bf
Rydberg atoms are ideal tools for quantum technologies \cite{adams_rydberg_2019}. Due to their large size, their dipole-dipole interaction at micrometer-scale distances and their coupling to external fields are huge. Recent experiments vividly exhibit their interest for quantum simulation \cite{barredo_atom-by-atom_2016,Bernien_Probing_2017,signoles_glassy_2021}, in spite of limitations due to the relatively short lifetime of optically-accessible Rydberg levels. These limitations motivate a renewed interest for the long-lived circular Rydberg states \cite{nguyen_towards_2018,Cohen_Quantum_2021}. However, detecting them is so far either destructive \cite{nguyen_towards_2018} or complex \cite{Cohen_Quantum_2021}. Moreover, alkali circular states can be manipulated only by microwave fields, unable to address individual atoms. Alkaline earth circular states, with their optically active second valence electron, can circumvent these problems. Here we show how to use the electrostatic coupling between the two valence electrons of strontium to coherently manipulate a circular Rydberg state with optical pulses. We also use this coupling to map the state of the Rydberg electron onto that of the ionic core. This experiment opens the way to a state-selective spatially-resolved non-destructive detection of the circular states and, beyond, to the realization of a hybrid optical-microwave platform for quantum technology.

\rm

Quantum simulation is one of the most advanced quantum technologies with impressive advances in the understanding of quantum many-body phenomena~\cite{altman_quantum_2021}. Atoms in optical lattices \cite{bakr_quantum_2009,cheuk_quantum-gas_2015}, ion traps \cite{blatt_quantum_2012,bohnet_quantum_2016,zhang_observation_2017} and Rydberg atoms \cite{barredo_atom-by-atom_2016,Bernien_Probing_2017,scholl_quantum_2021,ebadi_quantum_2021} are among the most promising platforms. The latter can be arranged in arbitrary three-dimensional arrays \cite{barredo_synthetic_2018}. They interact strongly through a short-range dipole-dipole interaction. Simulations involving up to 200 atoms have been performed \cite{scholl_quantum_2021,ebadi_quantum_2021}. However, all experiments so far involve low-angular-momentum laser-accessible Rydberg states. Their lifetime is limited by optical spontaneous emission to a few hundred microseconds, resulting in a maximum simulation time of a few microseconds for the largest arrays.

Circular Rydberg atoms \cite{hulet_rydberg_1983} do not suffer from this limitation. They have the maximum allowed magnetic quantum number $m = \ell = n-1$, where $n$ is the principal quantum number and $\ell$ the orbital angular momentum. Their lifetime is about 30 ms for $n\sim50$ and can even be enhanced up to a minute by controlling their microwave spontaneous emission \cite{nguyen_towards_2018}. Their mutual interaction is as strong as that of standard Rydberg atoms and is fully tailorable by applying controlled electric, magnetic and microwave fields \cite{nguyen_towards_2018}. However, alkali circular Rydberg atoms used so far are impervious to optical light and cannot be addressed individually by laser beams. 

In contrast, alkaline earth Rydberg atoms have an optically active ionic core. Their interest has been demonstrated in recent experiments using low-angular-momentum states, with dipole optical traps based on the core polarizability  \cite{bounds_Rydberg-Dressed_2018,wilson_trapped_2019} and detection schemes based on the immediate auto-ionization following the core excitation \cite{Lochead_number-resolved_2013,Madjarov_high-fidelity_2020}. As the auto-ionization rate decreases rapidly with $\ell$ \cite{McQuillen_Imaging_2013}, it becomes negligible for circular states \cite{Teixeira_Preparation_2020}. This opens the way to versatile optical manipulation of alkaline earth circular states.

Here, we show that the electrostatic coupling between the Rydberg electron of a strontium circular atom and the ionic core electron promoted to a metastable state induces an energy shift that depends on the state of both electrons. This shift allows us to condition the state of the ionic core electron to the value of $n$. Conversely, we use a laser pulse to coherently manipulate the state of the circular Rydberg electron, opening the way to a quantum interface between optical and microwave realms. 

\begin{figure}
 \centering
 \includegraphics[width=.99\linewidth]{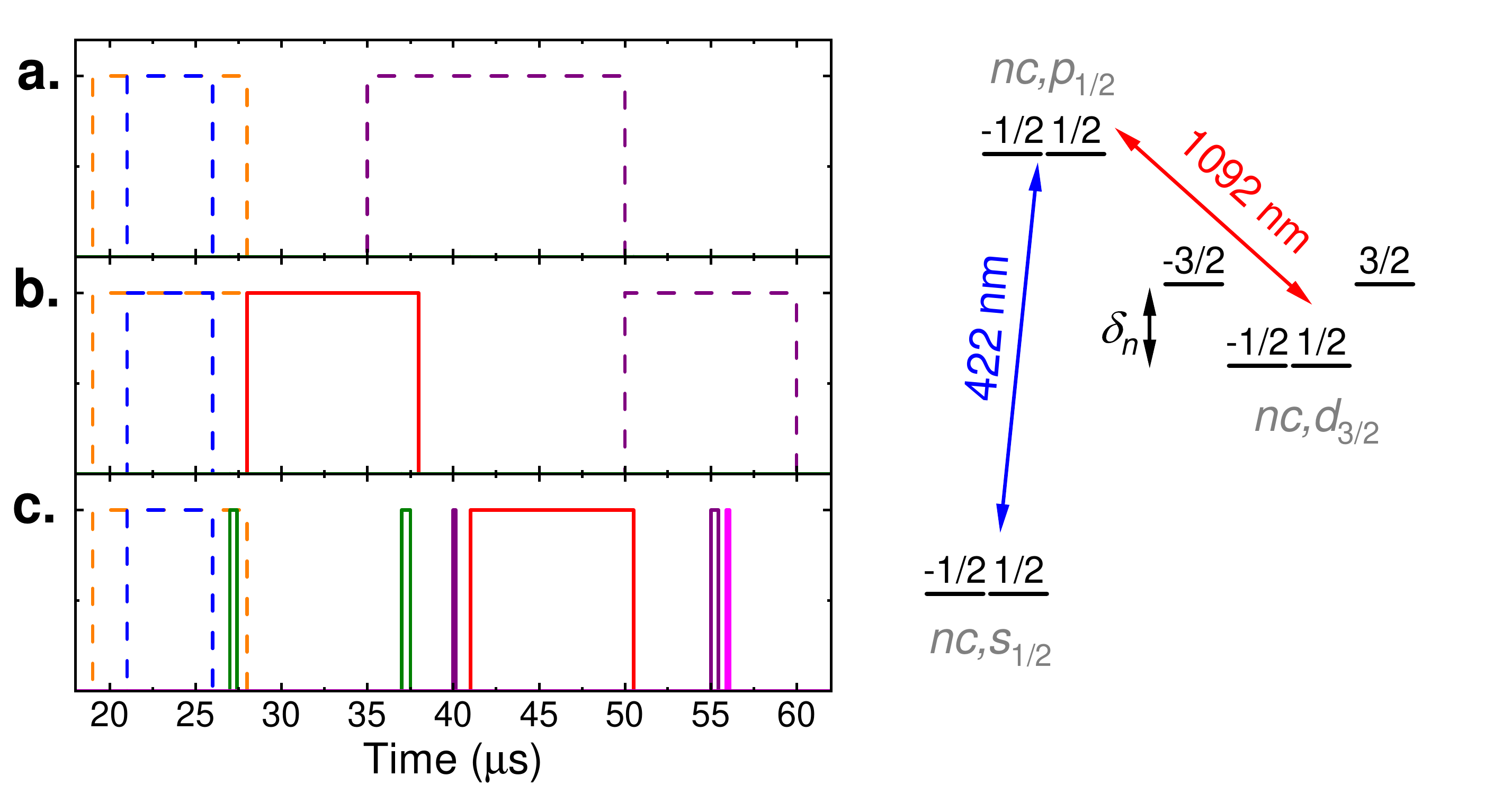}
 \caption{\bf Experimental sequences. \rm \changes{Laser and MW pulses applied after the circular state preparation. The dashed-blue, dashed-orange and solid-red rectangles show the 422 nm laser pulse, the resonant 1092 nm repumper pulse and the Raman pulses respectively. The solid-green, dashed-purple and solid-magenta rectangles represent the microwave pulses resonant with the $51c\rightarrow50c$, $51c\rightarrow49c$ and $51c\rightarrow53c$ transitions respectively. Note that the dashed rectangles represent pulses for which the atom-field interaction time is limited by the transit time of the atom through the corresponding laser or microwave mode and is thus shorter than the pulse duration.} (a) Microwave spectroscopy~: the \changes{optical pumping pulses} are followed by a 15 $\mu$s-long microwave pulse resolving the differential shift induced by the quadrupole effect. (b) Laser spectroscopy of $nc,4d_{3/2},|m_j|=3/2$ : the 422 nm optical pumping pulse takes place while a $\pi$-polarized repumper 1092 nm laser is on. We then apply the 1092 nm Raman pulses, followed by an ionic-core state-selective microwave pulse. (c) Optical switch experiment: after removing the residual $51c,5s_{1/2}$ atoms (green pulses) we apply two microwave $\pi/2$ pulses (purple) before and after the Raman pulse. \changes{A final $51c\rightarrow 53c$ pulse (magenta) is used to reduce the background due to the imperfect circular state preparation (see Methods). The left panel} shows the level structure of the ionic core. \changes{The numbers above the level represent the value of $m_j$.}}
 \label{fig:1}
\end{figure}

The circular Rydberg states are produced from a strontium thermal beam in a cryogenic environment. The atom first interacts with a set of three lasers followed by a circularization process involving a microwave pulse and a $\sigma^+$-polarized radio-frequency field (for details see \cite{Teixeira_Preparation_2020}). This process prepares the atom in the $51c,5s_{1/2}$ state, in which the first electron is in the circular Rydberg state with $n=51$ and the second in the ground state of the ionic core. The quantization axis is defined by a static electric field, set to 1.4 V/cm after circularization. Once the Rydberg electron is in the circular state, the ionic core level structure is very close to that of the Sr$^+$ ion [right panel of Fig.~\ref{fig:1}]. A 422~nm laser pulse can excite the atom into the short-lived $51c,5p_{1/2}$ level. It decays both into $51c,5s_{1/2}$ and into the metastable $51c,4d_{3/2}$ state (branching ratio 17:1). An atom interacting with a resonant 422 nm laser is thus quickly optically pumped into a statistical mixture of the $m_{j}$ sublevels of $51c,4d_{3/2}$.  A $\pi$-polarized repumper beam at 1092~nm, resonant on the $4d_{3/2}$ to $5p_{1/2}$ ionic core transition, can be used to selectively empty the $m_j=\pm 1/2$ sublevels. At the end of the experimental sequence, we finally measure the state of the Rydberg electron in a state-selective field-ionization detector \cite{Teixeira_Preparation_2020}.

The $nc,5s_{1/2}$ and $nc,4d_{3/2}$ states have different electronic charge distributions of the ionic core. Unlike the $5s_{1/2}$ state, the $4d_{3/2}$ state has a static quadrupole moment. Its interaction with the electric field created by the Rydberg electron leads to a shift, $V_{m_j}$, of the $nc,4d_{3/2},m_j$ levels given by~\cite{itano_external-field_2000} 
\begin{equation}
V_{m_j}= 
\left\{
\begin{array}{c}
   h\delta_n/2  \mbox{  if  } |m_j|=3/2 \\
  - h\delta_n/2  \mbox{  if  } |m_j|=1/2  
\end{array}\right.
\label{eq-shiftabs}
\end{equation}
where $ {\delta_n}  = \dfrac {1} {h} \left|\dfrac{\partial E_z}{\partial z} \right |\Theta_{4d_{3/2}}$. Here, $\Theta_{4d_{3/2}}$ is the $4d_{3/2}$  quadrupole moment and $E_z$ is the electric field created by the Rydberg electron along the quantization axis $Oz$ at the position of the ionic core (see Methods). Since the field gradient ${\partial E_z}/{\partial z}$ is proportional to the inverse cube of the circular level orbit size, $\delta_n$ scales like $n^{-6}$. From the theoretical value of $\Theta_{4d_{3/2}}= 2.029(12)$ \cite{jiang_electric_2008}, we expect $\delta_{51}= 759$ kHz. This effect can be viewed either as a shift of the Rydberg levels due to the ionic core or as a shift of the ionic core levels due to the Rydberg electron.

\begin{figure}
 \centering
 \includegraphics[width=\linewidth]{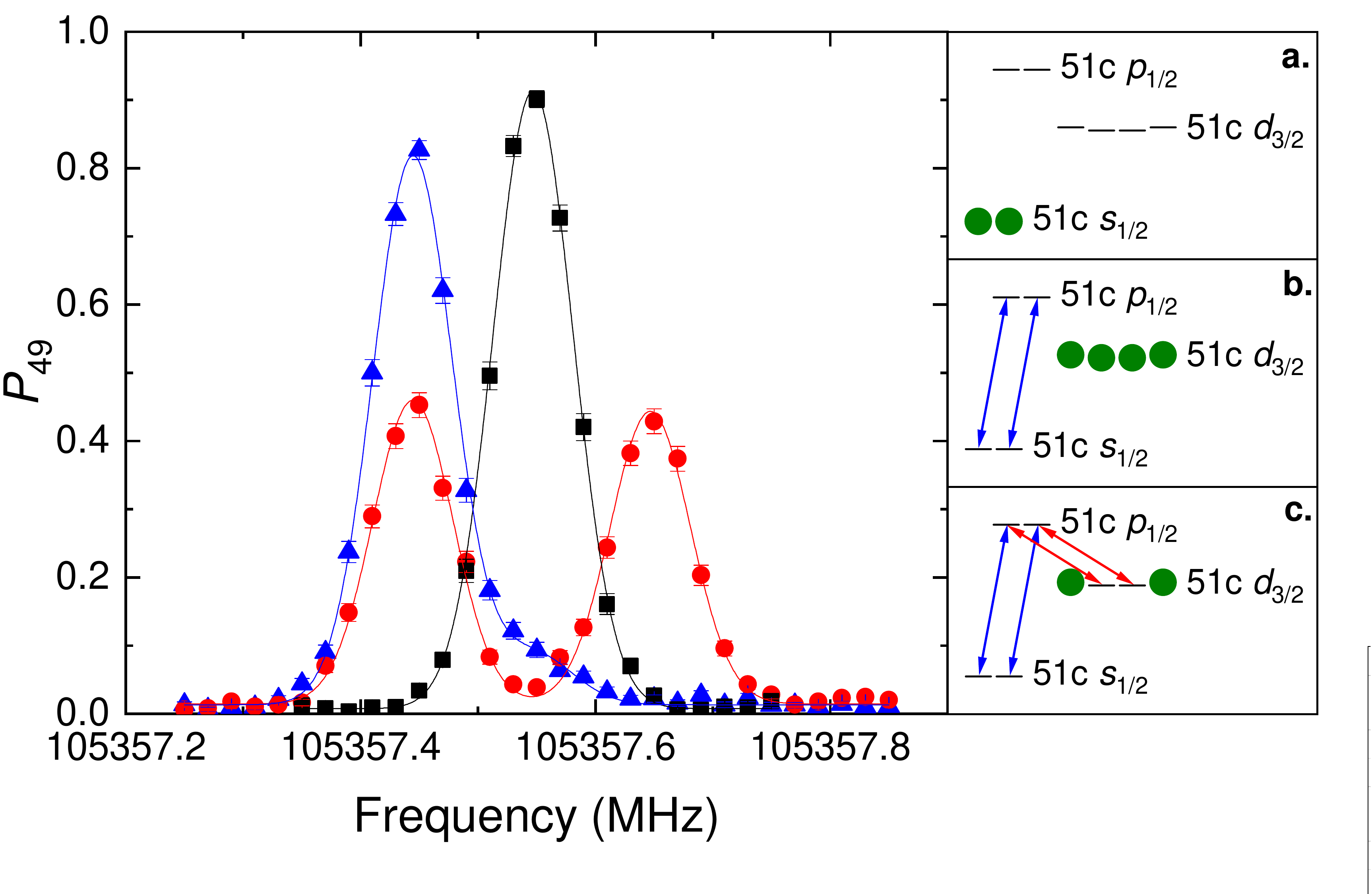}
 \caption{\bf Microwave spectroscopy. \rm Left:
 Microwave spectrum of the  $51c\rightarrow 49c$ transition without applying the 422 nm laser [black, inset (a)], after applying the 422 nm [red, inset (b)], after applying simultaneously the 422 nm laser and a $\pi$-polarized 1092 nm  repumper laser [blue, inset (c)]. The lines are gaussian fits. Right : black lines represent the different $m_j$ levels, red and blue arrows the 1092 nm and 422 nm lasers respectively, and green circles show the state to which the atomic population is optically pumped.}
 \label{fig:oreilles}
\end{figure}

As a first step, we investigate the level shift of the Rydberg electron by microwave spectroscopy [timing in Fig. \ref{fig:1}(a)]. We first measure the $51c,5s_{1/2}\rightarrow49c,5s_{1/2}$ two-photon transition frequency. We apply on the atom in the $51c,5s_{1/2}$ state a 15 $\mu$s-long microwave pulse and record the transfer probability to the $n=49$ manifold as a function of $\nu$, where $\nu/2$ is the microwave frequency (black points on Fig. \ref{fig:oreilles}). We observe a single resonance at $\nu=\nu_0 = 105.357546$ GHz. We then repeat the same experiment when the core electron is optically pumped by a 422 nm laser pulse in a statistical mixture of the $51c,4d_{3/2},m_j$ sublevels (red points on Fig. \ref{fig:oreilles}). The signal now  shows two resonant peaks  with frequencies $\nu_{3/2}$ and $\nu_{1/2}$, symmetric with respect to $\nu_0$, corresponding to the transitions $51c,4d_{3/2},|m_j|=3/2\rightarrow 49c,4d_{3/2},|m_j|=3/2$ and $51c,4d_{3/2},|m_j|=1/2\rightarrow 49c,4d_{3/2},|m_j|=1/2$ respectively. The splitting between these resonances, 202(2) kHz, measures the value of $|\delta_{49}-\delta_{51}|$. The heights of the red peaks are similar, and half of that of the black one. This is consistent with the expected equal population in all the $4d_{3/2},m_j$ sublevels. This equal balance can be altered by shining a $\pi$-polarized 1092~nm repumper laser during the 422 nm pulse. The atoms are then optically pumped into the $51c,4d_{3/2},|m_j|=3/2$ levels. Fig.~\ref{fig:oreilles} presents the corresponding microwave spectrum (blue points). From a fit, we estimate that 90~\% of the population is in $51c,4d_{3/2},|m_j|=3/2$. The remaining $\sim 10$~\% end up in $5s_{1/2}$ (see Methods).

\begin{figure}
 \centering
  \includegraphics[width=.9\linewidth]{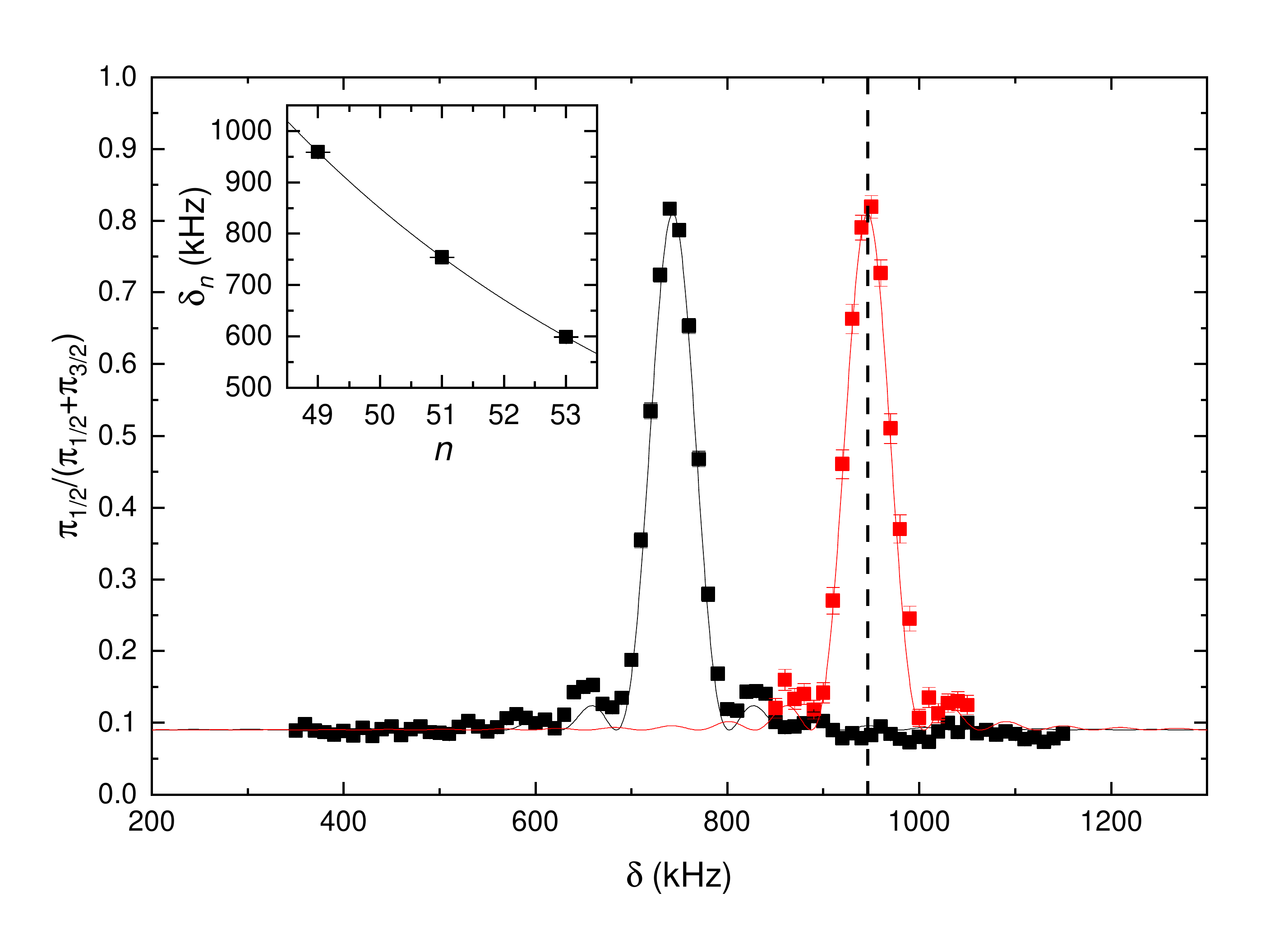}
 \caption{\bf Laser spectroscopy. \rm Raman transfer rate, $\pi_{1/2}/(\pi_{3/2}+\pi_{1/2})$, as a function of the detuning $\delta/2\pi$ between the two Raman beams when the atom is initially in the $51c,4d_{3/2},|m_j|=3/2$ state (black) or  $49c,4d_{3/2},|m_j|=3/2$ state  (red). The points are experimental with statistical error bar, the solid lines are the expected sinc function for a 17 $\mu$s pulse (the amplitude and offset of the sinc function have been adjusted). The vertical dashed line shows the optimal Raman frequency for a selective excitation of the ionic core when the Rydberg electron is in the $49c$ state.  Inset : $\delta_n$ as a function of $n$. The points are experimental (with the fit error bars smaller than the point size), the line is a fit to the $1/n^6$ law. }
 \label{fig:pipulse}
\end{figure}

We now investigate the influence of the Rydberg electron on the core levels [timing in Fig. \ref{fig:1}(b)]. We perform a Raman laser spectroscopy experiment between the $4d_{3/2},|m_j|=3/2$ and $4d_{3/2},|m_j|=1/2$ sublevels of the ionic core.  We prepare the atom in the $51c,4d_{3/2},|m_j|=3/2$ states by the optical pumping sequence described above. We then apply simultaneously $\pi$-polarized  and  $\sigma$-polarized 1092 nm laser pulses, respectively detuned from the $51c,4d_{3/2}\rightarrow51c,5p_{1/2}$ transition by $\Delta$ and $\Delta+\delta$ (\change{$\Delta\sim0.65$ GHz} and $\delta\ll\Delta$). We expect a resonant transfer from $|m_j|=3/2$ to $|m_j|=1/2$ for $\delta = \delta_{51}$. We probe the final populations  $\pi_{3/2}$ and $\pi_{1/2}$ of the $|m_j|=3/2$ and $|m_j|=1/2$ levels \changes{with $m_j$-selective microwave probe pulses with frequencies $\nu_{3/2}$ and $\nu_{1/2}$ respectively.}  Fig.~\ref{fig:pipulse} shows the Raman transfer rate, $\pi_{1/2}/(\pi_{3/2}+\pi_{1/2})$, as a function of $\delta$ for a 17~$\mu$s-long Raman $\pi$ pulse (black points) and compares it to the results of \change{the same experiment when the atom is prepared in $49c$ at the beginning of the sequence instead.} 
The two resonances are clearly separated, showing that it is possible to selectively transfer the ionic core electron if and only if the Rydberg electron is in the $n=49$ circular state, by choosing a Raman pulse frequency corresponding to the maximum of the red curve (dashed line on Fig. 3).

We have measured $\delta_n$ for $n=53,51$ and 49 by recording similar spectra. \changes{We get rid of the light shift induced by the Raman lasers by recording spectra for different laser powers and extrapolating to zero intensity} (see Methods). The inset in Fig. \ref{fig:pipulse} shows the resulting values of $\delta_n$. The data is in excellent agreement with the expected $n^{-6}$ law. A fit provides the first measurement of the $4d_{3/2}$ quadrupole moment $\Theta_{4d_{3/2}} = 2.025(3)$ atomic units, in \changes{perfect} agreement with the theoretical predictions \cite{jiang_electric_2008}. 

\begin{figure}
 \centering
 \includegraphics[width=\linewidth]{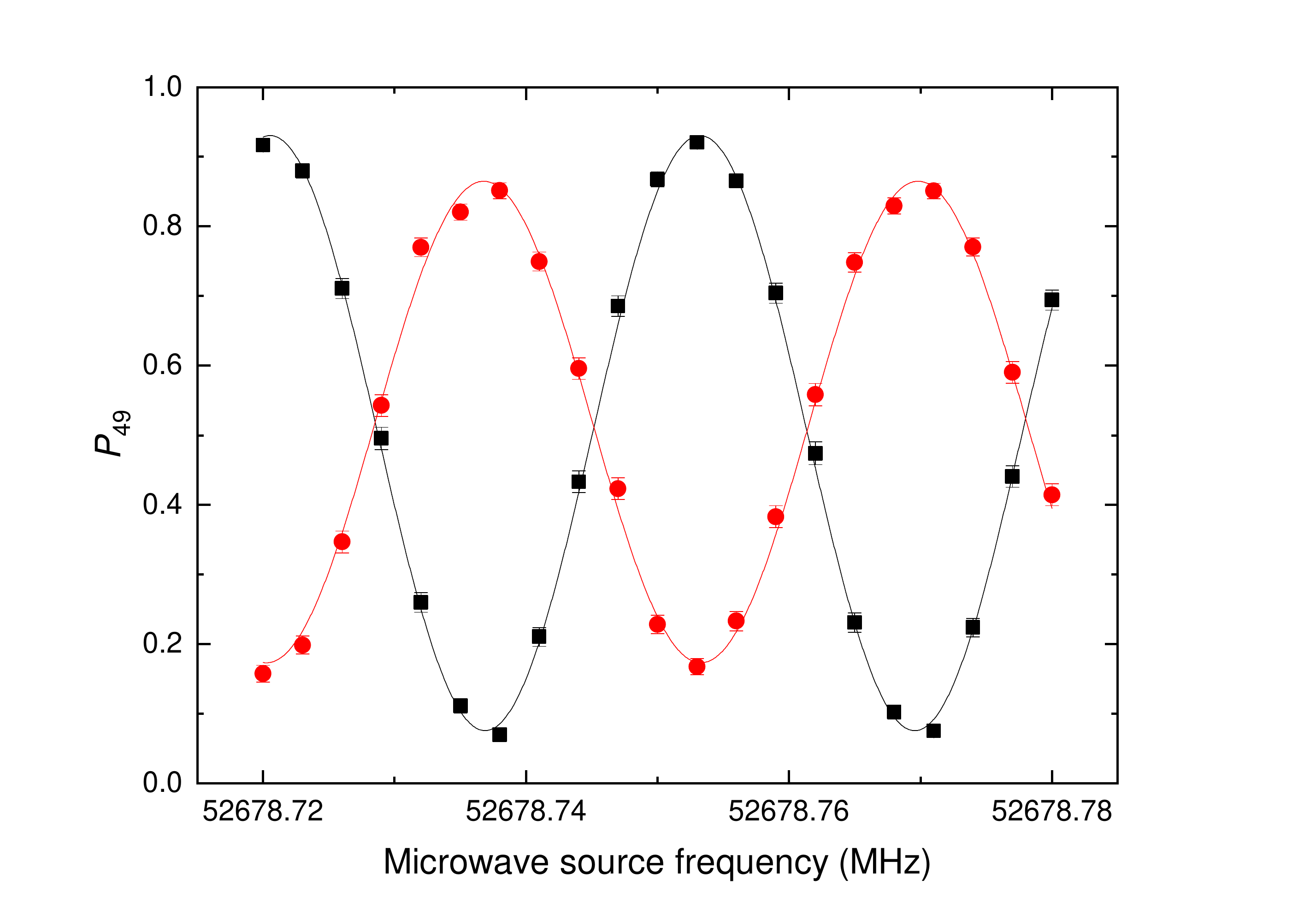}
 \caption{\bf Ramsey fringes \rm. Probability to detect the atom in the $49c$ state at the end of the Ramsey sequence with (red) or without (black) the 1092 nm Raman pulses as a function of the frequency of the microwave source. The points are experimental (with statistical error bars), the line a sine fit to the data. }
 \label{fig:2pipulse}
\end{figure}

We now show that we can use the Raman pulse applied on the ionic core to coherently manipulate the Rydberg electron by flipping the phase of a coherent superposition of two circular states [experimental timing in Fig. \ref{fig:1}(c)]. We start from an atom optically pumped into the $51c,4d_{3/2},|m_j|=3/2$ sublevels. We remove the residual population of $51c,5s_{1/2}$ level by selectively transferring it to the $n=50$ manifold using \changes{a pair of microwave pulses} (see Methods). We implement a Ramsey interferometer by applying two short $\pi/2$ microwave pulses (duration $0.15$ and $0.45$ $\mu$s, the difference being due to the variation of the microwave field local amplitude as the atoms move inside the cryostat), short enough not to resolve the quadrupole-induced shift, separated by 15 $\mu$s on the $51c\rightarrow49c$ transition. They prepare and analyse a coherent superposition of these two circular Rydberg states. Fig. \ref{fig:2pipulse} (black dots) presents the probability to detect the atom in the $n=49$ manifold as a function of the frequency of the microwave source and, hence, as a function of the relative phase of the microwave pulses.

We now flip the phase of the circular state coherent superposition in the time interval between the microwave pulses by applying a 9.5 $\mu$s-long $2\pi$ Raman pulse tuned on the $49c,4d_{3/2},|m_j|=3/2\rightarrow 49c,4d_{3/2},|m_j|=1/2$ transition. Ideally this pulse should imprint a phase shift of $\pi$ on the $49c,4d_{3/2}$ component of the superposition without affecting the $51c,4d_{3/2}$ component. However, as its duration is not much larger than $1/(\delta_{49}-\delta_{51})$, it  induces a spurious off-resonant Rabi oscillations on the $51c,4d_{3/2},m_j=\pm3/2\rightarrow 51c,4d_{3/2},m_j=\pm1/2$ transition. In order to minimize this effect, we optimize the detuning $\delta$ and the power of the pulse to induce an integer number of these off-resonant Rabi oscillations (see Methods).
 
Fig. \ref{fig:2pipulse} (red dots) presents the Ramsey fringe pattern in the presence of the Raman pulse. 
The expected  $\pi$ phase-shift of the Ramsey fringes is conspicuous.
The contrast of the fringes is slightly reduced by the Raman pulse, due to the spurious excitation of the $49c,5p_{1/2}$ level, followed by an incoherent spontaneous emission to $49c,5s_{1/2}$. This transfer could be reduced by increasing the Raman detuning $\Delta$. 
In this experiment, the Raman pulse behaves as an optical switch controlling the final state of the Rydberg atom. A control of the sign of $m_j$, which is not available in the present geometry, would directly provide a quantum gate between an optical transition of an ion and a Rydberg qubit.

We have demonstrated that the electrostatic coupling between the two valence electrons of an alkaline-earth circular Rydberg atom leads to an optical manipulation of the core electron conditioned to the state of the Rydberg one.  We have also shown that the optical excitation of the core coherently controls the state of the Rydberg electron. 

These two effects together open fascinating perspectives for quantum simulation and quantum information with Rydberg atoms. The first prepares the ionic core electron into a metastable level depending on the value of $n$. Combined with the optical shelving techniques developed for ion trap experiments \cite{Leibfried_quantum_2003}, it opens the way to controlling the fluorescence of the ionic core by the state of the Rydberg electron. This fluorescent emission can be easily detected for individual laser-cooled atoms \cite{McQuillen_Imaging_2013}. It would then be possible to detect a single trapped circular Rydberg atom in a quantum non-destructive and state-selective way.
For circular atoms individually trapped in optical tweezer, the second  effect makes it possible to individually address a circular atom in an array with tightly focused laser beams, an essential tool for quantum simulation and quantum information, lacking with the usual microwave addressing of circular states.  

The methods presented here equally apply to all alkaline-earth-like atoms~\cite{Madjarov_high-fidelity_2020,Cohen_Quantum_2021,lehec_isolated_2021,Trautmann_spectroscopy_2021}. Our experiment shows that alkaline earth circular Rydberg states have the potential to combine the best features of the fluorescence detection and shelving techniques of ion traps with those of the Rydberg atoms. It opens a link between optical and microwave quantum information platforms. It removes crucial bottlenecks on the road towards an extensive use of circular Rydberg atoms in a variety of quantum technologies.

\section{Methods}

\subsection{Calculation of quadrupole shift, quantum defect, off-resonant dipole}

To calculate the energy levels of the strontium atom when the outer electron is in the circular state $nc$, we consider the two active electron (TAE) Hamiltonian 
$$  H = \frac{p_1^2} 2 + \frac{p_2^2} 2  + V_{\ell_1}(r_1) + V_{\ell_2}(r_2) + \frac 1 {|\mathbf r_1 - \mathbf r_2|}  $$
written using atomic units and \changes{where index 1 and 2 respectively refer to the Rydberg and core valence electron. The $V_{\ell_i}(r_i)$ term} is an angular-momentum-dependent semi-empirical model potential representing the Sr$^{2+}$ ion  \cite{ye_production_2013}. The term $V_{12} = 1/ {|\mathbf r_1 - \mathbf r_2|}$ can be expanded using Laplace expansion $$V_{12} = \sum_{k = 0}^\infty W_k$$ where 
$$W_k= \sum_{m = -k}^k (-1)^m Q^{(k)}_{m}C^{(k)}_{-m}$$
with 
$$  Q^{(k)}_{m} =  -\sqrt{  \frac{4 \pi}{2k+1}} r_2^k Y_{k,m}(\theta_2,\varphi_2)$$
and
$$ C^{(k)}_{m} =-\sqrt{  \frac{4 \pi}{2k+1}} r_1^{-(k+1)}Y_{k,m}(\theta_1,\varphi_1).$$
We have assumed $r_1 >r_2$. The first term $W_0 = 1/r_1$ corresponds to the shielding of the Sr$^{2+}$ ionic core charge by the second electron and only depends on $r_1$. \changes{Neglecting multipolar interaction terms with $k>2$}, one can rewrite $H$ as a sum of three terms $H = H_1 + H_2 + V$ where
$$  H_1 =\frac{p_1^2} 2 + V_{\ell_1}(r_1)  + W_0 $$
$$ H_2 =  \frac{p_2^2} 2  + V_{\ell_2}(r_2)  $$
$$ V =  W_1+ W_2  .$$

$H_1$ is the Hamiltonian of the Rydberg electron. Since $\ell_1\gg 1$, the Rydberg electron always remains far away from the ionic core and $V_{\ell_1}(r_1)  + W_0 \approx -1/r_1$. The eigenstate of $H_1$ are up to an excellent approximation 
the Rydberg levels $\ket R$ of the hydrogen atom. 

\change{$H_2$ is, up to a good approximation, the Hamiltonian of the valence electron of a Sr$^+$ ion, whose eigenvectors  are the strontium ion states $\ket {n_2\ell_{2}j_2}$.}

The final term, $V$ describes the residual coupling between the two valence electrons \changes{due to electric dipole  $W_1$ and quadrupole $W_2$ interactions.} It is much smaller than \changes{$H_1$ and $H_2$. We determine the effect of $V$ on the eigenstates states $\ket R \otimes \ket {n_2\ell_{2}j_2}$ of $H_1+H_2$  by a perturbative approach.}

\paragraph{First-order approximation.}

The static electric field of about 1.4 V/cm that defines the quantization axis is large enough to lift the degeneracy between the circular state and the other levels of the $n^{th}$ manifold. For the Rydberg electron in a circular state, we thus find the level energies to first order in $V$ by diagonalizing $V$ in the subspace of the four levels $\ket{nc, 4d_{3/2},m_j}$ ($m_j=-3/2,...,3/2$)

\changes{The matrix elements of the dipole operator $\bra{4d_{3/2},m_j}Q^{(1)}_{m}\ket{4d_{3/2},m_{j'}}$ are zero, and therefore $\langle W_1\rangle = 0$ in the considered subspace. }
We diagonalize $W_2$ by first calculating the matrix elements $\bra{nc} C^{(2)}_{m} \ket{nc}$. The only non-zero term corresponds to $m=0$
$$ \langle C^{(2)}_{0}\rangle = -\frac{1}{2}\left\langle  \frac {3 \cos^2\theta -1 } {r_1^{3}}\right\rangle = -\frac{1}{2}\left\langle  \frac {\partial E_z } {\partial z }\right\rangle $$
where $E_z$ is the $z$ component of the electric field created by the Rydberg electron on the ionic core.
As a result, $W_2 = Q^{(2)}_{0}C^{(2)}_{0}$, which is diagonal in the $\ket{nc, 4d_{3/2},m_j}$ basis.  
From the definition of $\Theta_{4d_{3/2}}$ \cite{itano_external-field_2000} we immediately get that $$\langle Q_{20}\rangle =  \Theta_{4d_{3/2}}$$ for $|m_j|=3/2$. Using angular momentum algebra we deduce  
$$\langle Q_{20}\rangle = - \Theta_{4d_{3/2}}$$  for $|m_j|=1/2$.

\paragraph{Higher-order terms.}

The second order shift induced by the $W_1$ term mostly originates from the coupling of $\ket{nc, 4d_{3/2},m_j}$ to the levels $\ket{R, 5p_{1/2}}$ and $\ket{R, 5p_{3/2}}$, \changes{where $R$ is a Rydberg state.}
It scales as $n^{-8}$ \cite{gallagher_radio_1982}. 
\change{A calculation based on Coulomb Green's function theory \cite{Poirier_Analysis_1998} provides a differential shift between $$\ket{51c, 4d_{3/2},m_j=\pm3/2}$$  and $$\ket{51c, 4d_{3/2},m_j=\pm1/2}$$ of $-2.7$ kHz (with an estimated uncertainty of $\pm1$ kHz), thus reducing $\delta_{51}$ by about 0.4 \%.}

The second order shift induced by the $W_2$ term mostly comes from one of the Stark levels with $m=n-3$, which is only 380 kHz away from the circular state. However, the coupling to this level by $W_2$ is small and the resulting energy shift is less than a kHz.

\subsection{Fit of \changes{microwave spectra}}
The fit of the spectral lines of Fig. \ref{fig:oreilles} is performed in three steps. We first fit the data without the 422 nm laser pulse (black dots) to a Gaussian line. We get the center frequency $\nu_0 = 105.3575470(6)$ GHz, the Gaussian width $w_0 =  78(1)$ kHz and the area $A_0$ of the peak. 

We then fit the data with the 422 nm laser (red dots) as a sum of two Gaussian peaks of fixed width $w_0$. We find the two resonance frequencies $\nu_{3/2} = 105.357444(1)$ GHz and $\nu_{1/2} = 105.357647(1)$ GHz. The two peak areas  $A_{3/2} = 0.49(2) A_0$ and $A_{1/2} = 0.47(2) A_0$ provide the relative population of the $|m_j|=3/2$ and $|m_j|=1/2$ sublevels. 

Finally, we fit the data with the 422 nm and $\pi$-polarized 1092 nm laser (blue dots). Experimentally, we set the 1092 nm pulse to be slightly longer than the 422 nm one. This ensures that the atoms remaining in any of the $|m_j|=1/2$ sublevels at the end of the optical pumping process are transferred back into $51c, 5s_{1/2}$. We thus expect the $|m_j|=1/2$ sublevels to be empty. We therefore fit the data as a sum of two Gaussian peaks of width $w_0$ and center frequencies $\nu_{3/2}$ and $\nu_{0}$. The result of the fit provides the two areas $A'_{3/2} = 0.89(2) A_0$ and $A'_{0} = 0.08(1) A_0$, showing that nearly 90$\%$ of the atoms are in the $|m_j|=3/2$ levels at the end of the laser pulses.

\changes{\subsection{Purification of $51c, 4d_{3/2}$}
For the Ramsey experiment, in order to improve the purity of $51c, 4d_{3/2}$, we first get rid of atoms left in $51c, 5s_{1/2}$ due to the imperfections of the optical pumping process. We apply after the optical pumping sequence} a set of two short ($\sim 0.5\,\mu$s) $\pi/2$ microwave-pulses, resonant with the $51c \rightarrow 50c$ transition and separated by 10 $\mu$s. This implements a Ramsey interferometer.  We set the frequency of the MW source so that an atom initially in $51c, 5s_{1/2}$ exits the interferometer in $50c, 5s_{1/2}$. The delay between the microwave pulses is chosen to be exactly $1/(\delta_{50}-\delta_{51})$. As a result, an atom initially in  $51c, 4d_{3/2}$ will exit the interferometer in $51c, 4d_{3/2}$. 
The two microwave pulses acts as an interference filter designed to transfer into the $n=50$ the $51c, 5s_{1/2}$ atom, while leaving the $51c, 4d_{3/2}$ atoms unaffected. Experimentally,  it decreases the population of $51c, 5s_{1/2}$ by a factor 20, while reducing the population of $51c, 4d_{3/2}, m_J = 3/2$ by only 6 \% (due to the experimental imperfections, see supplementary figure S1). This increases the $51c, 4d_{3/2}$ preparation purity to more than 99 \%.

\subsection{Extrapolation of $\delta_n$ and determination of $ \Theta_{4d_{3/2}}$}

In order to measure the absolute shift $\delta_n$, we initially prepare the atoms in the state  $nc, 4d_{3/2}, m_j=\pm 3/2$ ($n=49,51,53$) and record the probability to transfer the atom into the $ m_j=\pm 1/2$ sublevels as a function of the relative frequency of the two 1092 nm laser beams. Figure S2 shows the fitted resonance frequencies for different total powers of the 1092 nm light (keeping the power-ratio between the two polarization components approximatively constant). We clearly observe the light shift effect. We deduce the value of $\delta_n$ from the extrapolation at zero power of the data. 

\changes{The value of  $\delta_n$ can be expressed as 
$$\delta_n= B  \times \left(\frac{51}{n}\right)^6 + C \times \left(\frac{51}{n}\right)^8$$
where the $1/n^6$ term accounts for the first order shift induced by the quadrupole hamiltonian $W_2$ and the $1/n^8$ term accounts for the second order shift induced by the dipole hamiltonian $W_1$. However, the precision of our data does not allows us to directly fit $B$ and $C$. Instead, we fix the value of $C= -2.7 \pm 1 $~kHz to the calculated value mentioned above and fit the value of $B$ to the data presented in the inset of figure \ref{fig:oreilles}. We find $B= 757 \pm 1 \mbox{ kHz }$. The uncertainty essentially results from the uncertainty on the theoretical estimation of the $1/n^8$ dipole contribution to $\delta_n$.}
From this fit, we deduce the value of the quadrupole moment \change{$ \Theta_{4d_{3/2}} = 2.025(3)$~a.u}.

\subsection{Detection of the $51c$ atoms in the Ramsey interferometer experiment}

The circular state preparation has a finite efficiency. We estimate that $\sim 10\%$ of the atom that are detected at the ionization threshold of the $51c$ state are spurious non-circular high-angular momentum state \changes{of the same manifold} \cite{Teixeira_Preparation_2020} that are not resonant with the $51c\rightarrow49c$ $\pi/2$ microwave pulses. If we were to record the Ramsey fringes by directly detecting the number of atom at the $51c$ state ionization threshold, these atoms would add a constant background to the signal and reduce the fringes contrast. 

To ensure that we only detect the atoms that exit the interferometer in the $51c$ state, we apply after the second $\pi/2$ pulse a probe pulse that transfers the $51c$ atom in the $n=53$ manifold and detect the atom at the threshold of $53c$. The duration of the pulse ($0.2\, \mu$s) is chosen to be both long enough to ensure that only the circular states of the $n=51$ manifold are transferred and short enough to transfer the atom regardless of the state of the ionic core electron. 

\subsection{Off-resonant Rabi oscillations}

In order to selectively flip the phase of the $\ket{49c,4d_{3/2},m_j=\pm3/2} $ in the Ramsey experiment (Fig.  \ref{fig:2pipulse}) we apply a resonant $2\pi$ pulse on the $49c,4d_{3/2},|m_j|=3/2\rightarrow 49c,4d_{3/2},|m_j|=1/2$ transition. 
The effective Rabi frequency $\tilde\Omega/2\pi \sim 100$ kHz is not negligible compared to the detuning $\tilde\Delta= 0.2$ MHz between the $51c,4d_{3/2},|m_j|=3/2\rightarrow 51c,4d_{3/2},|m_j|=1/2$ and $49c,4d_{3/2},|m_j|=3/2\rightarrow 49c,4d_{3/2},|m_j|=1/2$ transitions. As a result, the laser pulse also induces a small off-resonant Rabi oscillation between $51c,4d_{3/2},|m_j|=3/2$ and $51c,4d_{3/2},|m_j|=1/2$ at an frequency $[(\tilde\Omega/2\pi)^2 + \tilde\Delta^2]^{1/2}$. 
We choose $\tilde\Omega$ such that $[(\tilde\Omega/2\pi)^2 + \tilde\Delta^2]^{1/2}= 2(\tilde\Omega/2\pi) $ (Fig. S3) to ensure that the atoms in both $51c$ and $49c$ end up in their initial $|m_j|=3/2$ sublevel at the end of the laser pulse.

Due to the small off-resonant Rabi oscillation, the part of the atomic wavefunction in the $n=51$ manifold also accumulates a phase due to the laser pulse. However, it is possible to compensate for this effect and ensure that the Ramsey fringes with and without laser are exactly out of phase by slightly adjusting $\delta$. We choose a detuning $\delta = \delta^*$ shifted by $\sim 20$ kHz from the $49c,4d_{3/2},|m_j|=3/2\rightarrow 49c,4d_{3/2},|m_j|=1/2$ resonance (see Supplementary Figure S4).

\change{
\section{Acknowledgement}

We thank I. Dotsenko for experimental support. 
Financial support from the Agence Nationale de la Recherche under the project ``SNOCAR'' (167754) is gratefully acknowledged. This publication has received funding from the European Union's Horizon 2020 under grant agreement No 786919 (Trenscrybe) and 765267 (QuSCo).

\section{Author contribution}
$^*$ These authors contributed equally.

A.M., L.L., A.C., R.C.T., J.M.R., M.B. and S.G. contributed to the experimental set-up. S.G. and M.P. performed the numerical simulations. A.M., L.L, A.C. and S.G. collected the data and analysed the results. M.B and S.G. jointly supervised the experiment. All authors discussed the results and the manuscript.

\section{Data availability}
The datasets generated during the current study are available from the corresponding author on reasonable request.

\section{Competing interest}
The authors declare no competing interests.}


%

\setcounter{figure}{0}
\renewcommand{\thefigure}{S\arabic{figure}}

\begin{figure*}
\sc{\huge{Supplementary figures}}
 \centering
 \includegraphics[width=.9\linewidth]{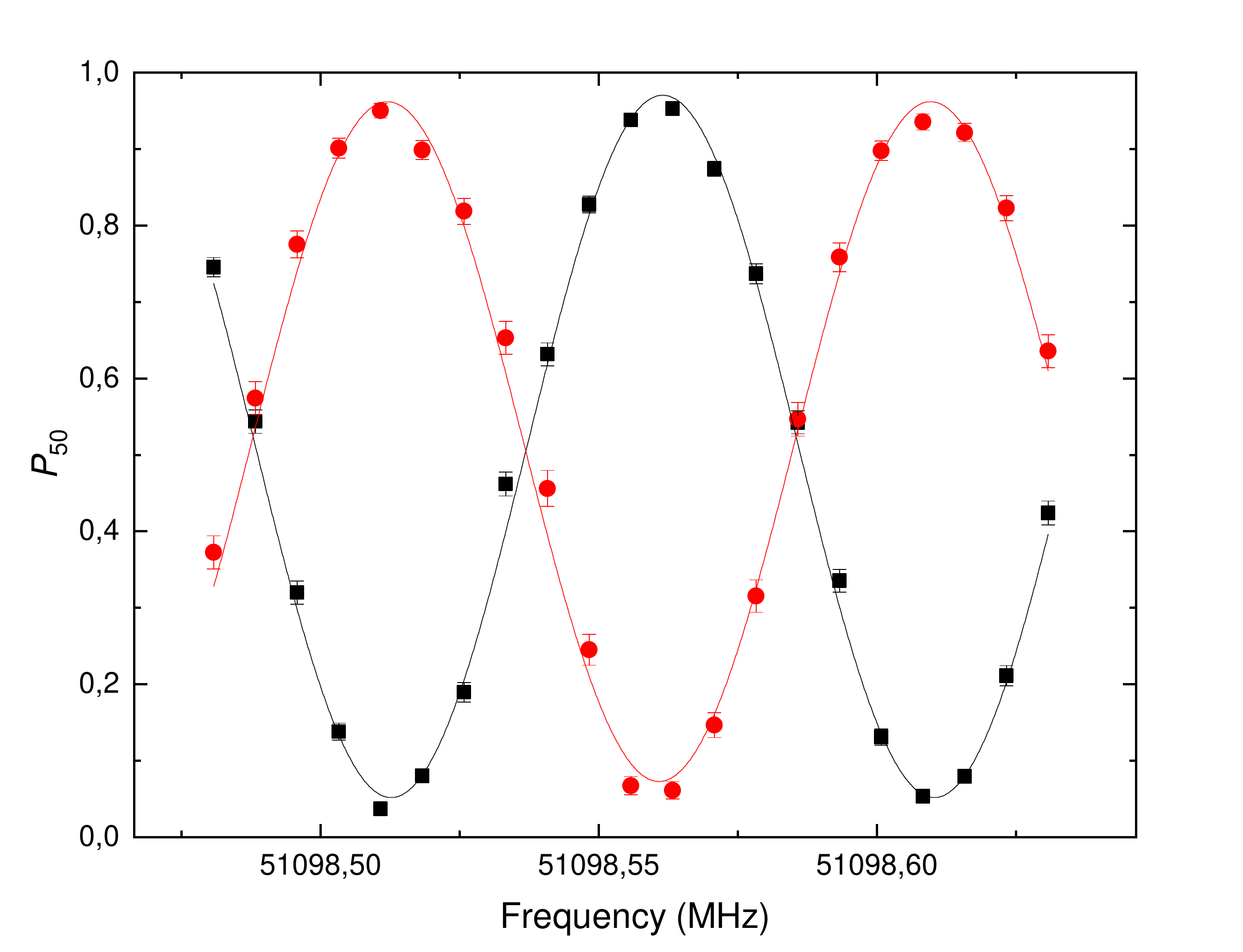}
 \caption{\bf Filtering of the $51c 5s_{1/2}$ atoms \rm. Ramsey fringes on the $50c-51c$ transition for an atom initially in the $51c 5s_{1/2}$ state (black) or in a mixture of $51c 4d_{3/2}$ sublevels (red). The points are experimental with statistical error bars, the solid lines are the result of fit to a sine. The spacing between the microwave $\pi/2$ pulse is chosen so that the fringes are exactly in phase opposition. The atoms in the state $51c$ are detected using a probe tuned on the $51c\rightarrow 53c$ transition. This ensures that non-circular $n=51$ states, present due to the imperfection of the preparation process, do not contaminate the signal. This enables us to better estimate the efficiency of the interference filter. For a frequency of 51.098516 GHz, 95 \% of the $51c 5s_{1/2}$ atoms are transferred to the $n=50$ manifold, while 94 \% of the $51c 4d_{3/2}$ remains in $n=51$. 
}
 \label{fig-filter}
\end{figure*}
 
\begin{figure*}

 \centering
 \includegraphics[width=.9\linewidth]{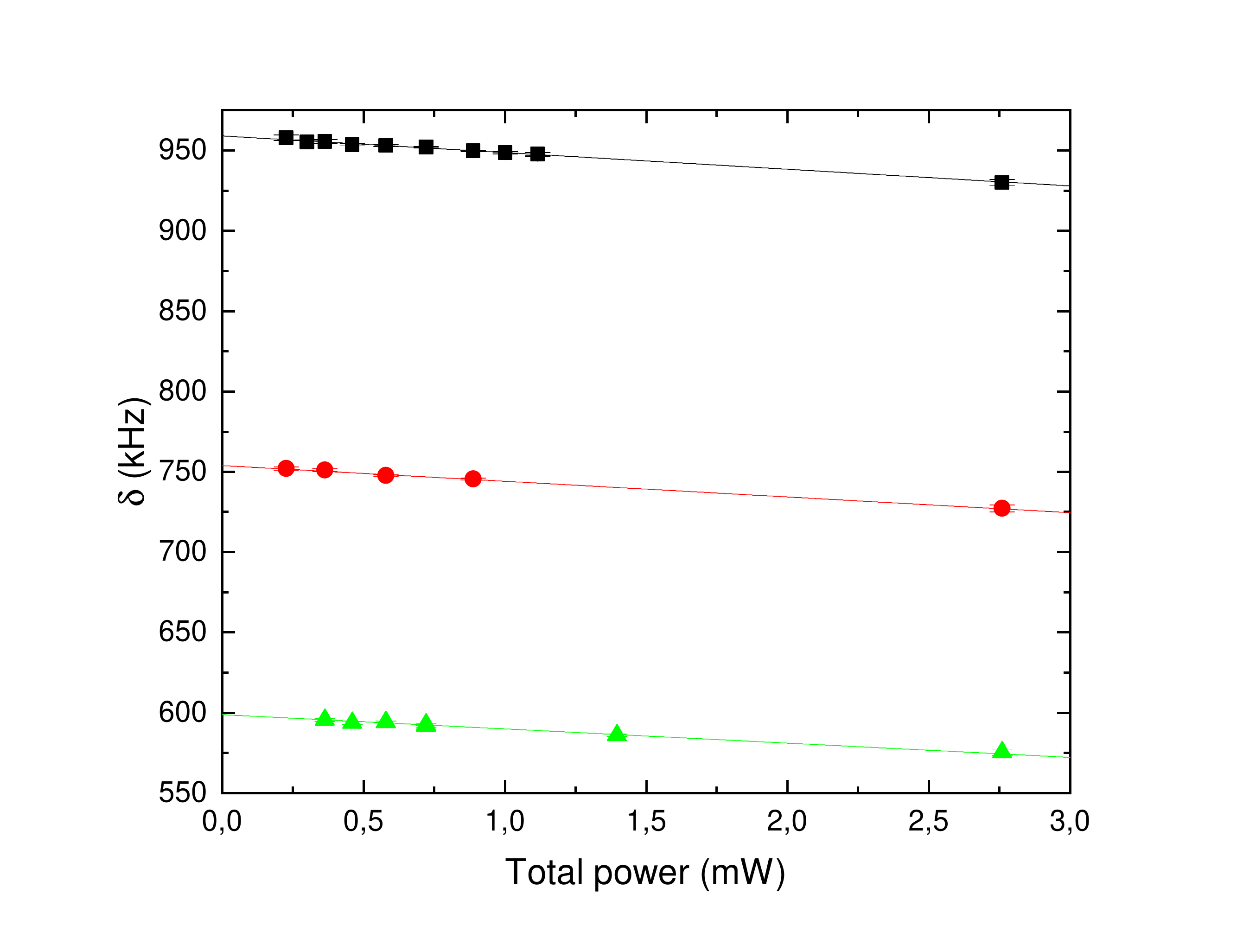}
 \caption{\bf Determination of $\delta_n$ \rm. Values of the Raman detuning $\delta$ inducing a resonant transfer from $nc,4d_{3/2},|m_j|=3/2$ to $nc,4d_{3/2},|m_j|=1/2$ for different powers of the 1092nm light and different values of $n$ [$n=49$ (black), $n=51$ (red) and $n=53$ (green)]. The points are experimental and the error bars represent the uncertainty of the fit. The ratio between the intensities of the $\pi$ and $\sigma$ components is 1.3:1, which induces a differential light shift proportional to the total intensity. To determine $\delta_n$, we thus linearly extrapolate the value of the resonant detuning at zero intensity (solid lines). 
}
 \label{fig-spectro1092}
\end{figure*}

\begin{figure*}
 \centering
 \includegraphics[width=\linewidth]{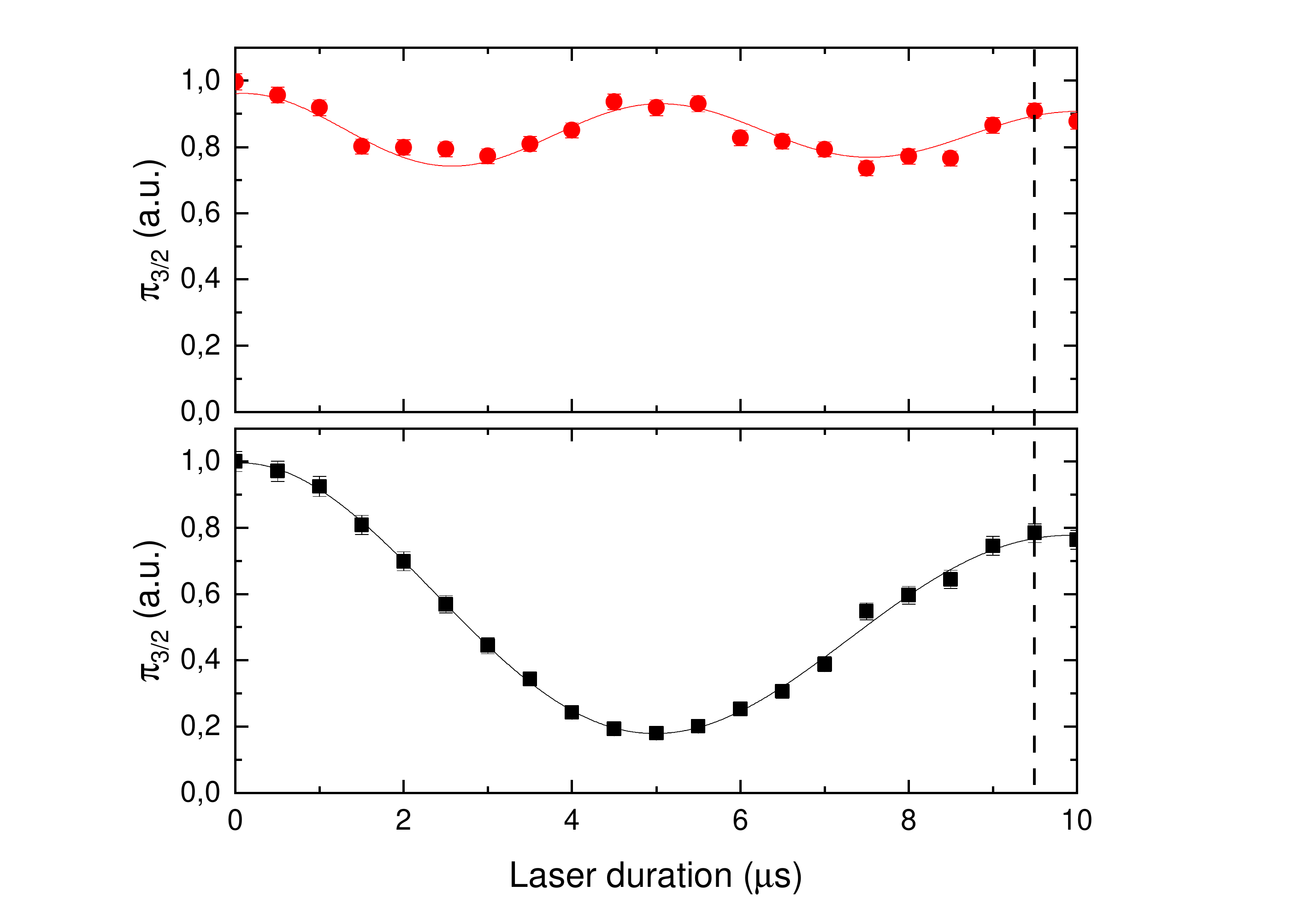}
 \caption{\bf Rabi oscillations induced by the Raman pulse\rm. Population $\pi_{3/2}$ of the $51c,4d_{3/2},|m_j|=3/2$ levels (upper pannel, red points) and the $49c,4d_{3/2},|m_j|=3/2$ levels (lower pannel, black points) as a function of the Raman laser pulse duration for a detuning $\delta = \delta^*$. The points are experimental, the line is a fit to the data using an exponentially damped sine function. The dashed line indicates the optimal pulse duration in the experiment.}
 \label{fig-2pi4pi}
\end{figure*}

\clearpage

\begin{figure*}
 \centering
 \includegraphics[width=.9\linewidth]{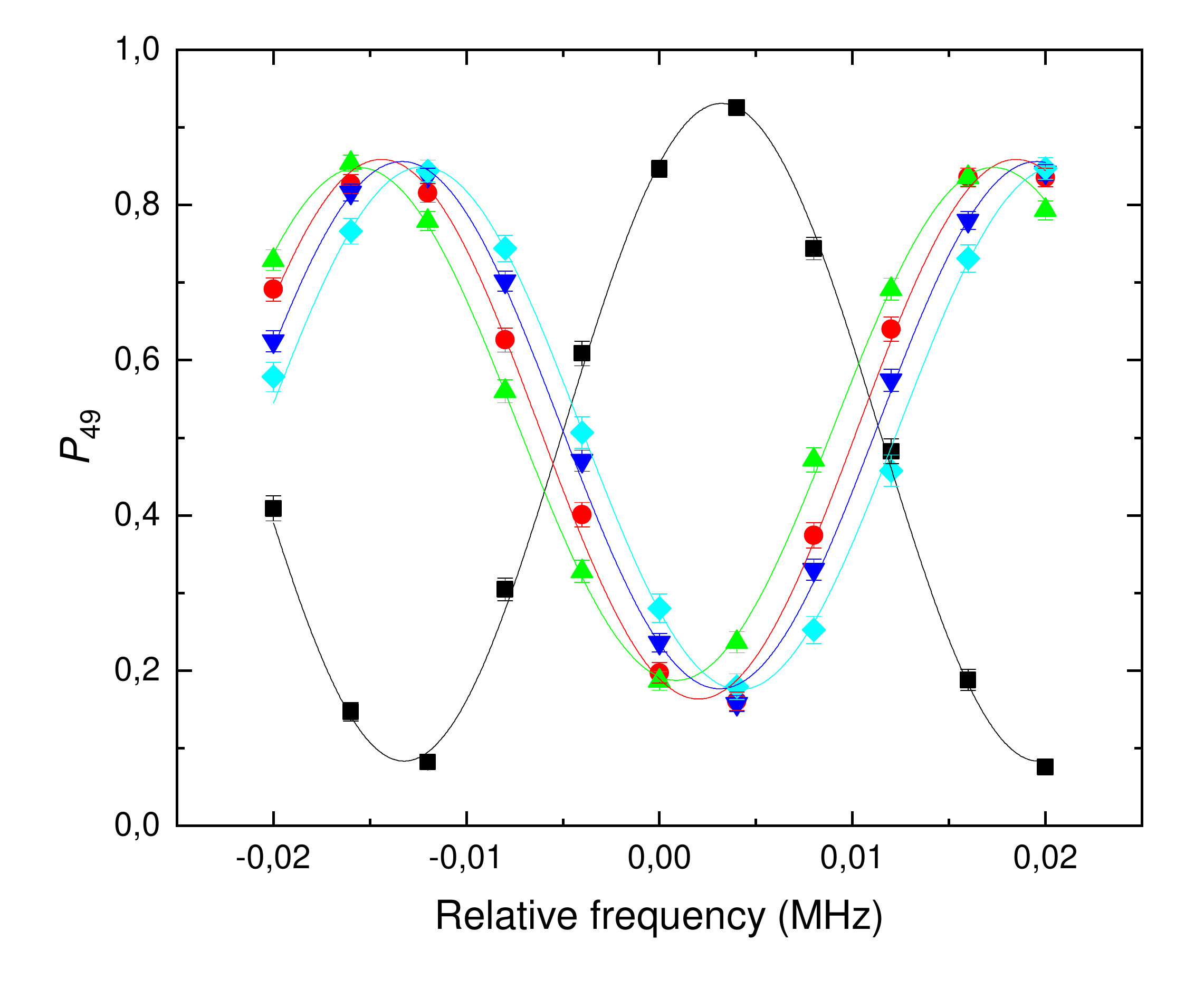}
 \caption{\bf Tuning of the phase shift induced by the Raman pulse\rm.
 Probability to detect the atom in the $49c$ state at the end of the Ramsey sequence with
 the 1092 nm Raman pulses for $\delta = 910$~kHz (green), $\delta = 900$~kHz (red), $\delta  = \delta^*= 890$~kHz (blue), $\delta = 880$~kHz (cyan) and without 1092 nm pulse (black). The points are experimental with statistical error bars, the lines are a sine fit to the data. For this laser power, the $49c,4d_{3/2},|m_j|=3/2\rightarrow 49c,4d_{3/2},|m_j|=1/2$ transition is resonant at $\delta = 913$~kHz.}
 \label{fig-deltastar}
\end{figure*}

\end{document}